\documentclass{article}

\usepackage{microtype}
\usepackage{graphicx}
\usepackage{booktabs} 
\usepackage{subcaption} 
\usepackage{float}
\usepackage{dirtree}
\usepackage{caption}  

\usepackage{hyperref}

\usepackage[accepted]{icml2025}

\usepackage{amsmath}
\usepackage{amssymb}
\usepackage{mathtools}
\usepackage{amsthm}

\usepackage[capitalize,noabbrev]{cleveref}

\theoremstyle{plain}

\theoremstyle{definition}

\theoremstyle{remark}

\usepackage[textsize=tiny]{todonotes}

\icmltitlerunning{}

\begin{document}

\twocolumn[
\icmltitle{Multimodal Modeling of CRISPR-Cas12 Activity Using Foundation Models and Chromatin Accessibility Data}



\icmlsetsymbol{equal}{*}

\begin{icmlauthorlist}

\icmlauthor{Azim  Dehghani Amirabad}{jj}
\icmlauthor{Yanfei Zhang}{jj}
\icmlauthor{Artem Moskalev}{jj}
\icmlauthor{Sowmya Rajesh}{jj}
\icmlauthor{Tommaso Mansi}{jj}
\icmlauthor{Shuwei Li}{jj}
\icmlauthor{Mangal Prakash*}{jj}
\icmlauthor{Rui Liao*}{jj}
\end{icmlauthorlist}


\icmlaffiliation{jj}{Johnson \& Johnson Innovative Medicine}

\icmlcorrespondingauthor{Azim Dehghani Amirabad}{azim.dehghani@gmail.com}
\icmlcorrespondingauthor{Mangal Prakash}{mpraka12@its.jnj.com}
\icmlcorrespondingauthor{Rui Liao}{rliao2@its.jnj.com}


\icmlkeywords{Machine Learning, ICML}

\vskip 0.3in
]



\printAffiliationsAndNotice{\icmlEqualContribution} 

\begin{abstract}
Predicting guide RNA (gRNA) activity is critical for effective CRISPR-Cas12 genome editing but remains challenging due to limited data, variation across protospacer adjacent motifs (PAMs—short sequence requirements for Cas binding), and reliance on large-scale training. We investigate whether pre-trained biological foundation model—originally trained on transcriptomic data—can improve gRNA activity estimation even without domain-specific pre-training. Using embeddings from existing RNA foundation model as input to lightweight regressor, we show substantial gains over traditional baselines. We also integrate chromatin accessibility data to capture regulatory context, improving performance further. Our results highlight the effectiveness of pre-trained foundation models and chromatin accessibility data for gRNA activity prediction.

\end{abstract}

\section{Introduction}
CRISPR-based genome editing enables programmable DNA modification, with Cas12 (Cpf1) offering advantages over the widely used Cas9 due to its distinct protospacer adjacent motif (PAMs—short sequence requirements for Cas binding) requirements and cleavage patterns \cite{zetsche2015cas12, kleinstiver2016cpf1}. A core challenge in CRISPR applications is predicting the activity of guide RNAs (gRNAs), short sequences that direct the Cas protein to specific genomic loci. Effective gRNA selection is critical for both editing efficiency and safety \cite{doench2016optimized}.

Existing models use handcrafted features or one-hot encoded sequences to predict gRNA activity, achieving modest performance \cite{hsu2013dna, xu2015comprehensive}. More recent deep learning approaches, including CNNs trained on large-scale screens, have improved prediction accuracy and incorporated additional features such as chromatin accessibility to account for regulatory context \cite{kim2018deep}. However, these models are typically sensitive to specific experimental settings, such as the variant of the genome-editing enzyme (Cas protein) used and the local sequence motifs required for binding (PAM sequence). As a result, they often fail to generalize across different enzyme variants or sequence constraints, limiting their applicability to broader genome engineering tasks.

In contrast, pre-trained biological Foundation Models (FMs) have shown strong generalization across a variety of genomic, transcriptomic, and proteomic tasks \cite{brandes2022nttransformer, chen2022rnafm, frazer2021deeper, gao2022unfolded, elnaggar2021prottrans}. These models produce rich, contextualized embeddings that capture structural and regulatory signals beyond local sequence patterns. However, their utility for CRISPR gRNA prediction—particularly for Cas12—remains unexplored.
Unlike DNA, RNA, and protein domains where pre-training such models is feasible due to abundant data, the gRNA space lacks sufficient data to support large-scale pre-training. This raises a critical question: Can we instead adapt existing DNA or RNA FMs to improve gRNA activity prediction, even though gRNAs are short (~30-50 nucleotides) and may lie outside the distribution existing off-the-shelf FMs were trained on? From a biological standpoint, this is a plausible strategy—gRNA sequences are functionally embedded in broader genomic and transcriptomic contexts, and their activity depends on features such as sequence composition, secondary structure, and chromatin accessibility that are captured in part by nucleotide-based FMs.

In this work, we evaluate two different FMs: RNA-FM~\citep{chen2022rnafm}, trained on transcriptomes, and DNABERT-2~\citep{zhou2024dnabert}, trained on genomic DNA. Using their embeddings as fixed input to lightweight CNN regressor, we evaluate their effectiveness for downstream gRNA on-target efficiency prediction. We incorporate chromatin accessibility profiles from ATAC-seq assay as an additional input modality, capturing the openness of genomic regions and their broader regulatory context. This epigenetic signal complements sequence information and can enhance prediction accuracy, even when using strong pretrained embeddings. To sum up, we make the following contributions.

\begin{itemize}
    \item We demonstrate that embeddings from a transcriptomic foundation model significantly outperform existing baselines in predicting CRISPR-Cas12 gRNAs, showcasing their ability to generalize to out-of-distribution sequences.
    \item We introduce a new dataset derived from public source curation, aligning ATAC-seq chromatin accessibility data to gRNA target loci. This structured epigenomic signal not only enhances predictions when integrated with sequence embeddings but also provides valuable resources for the community to advance research in this area. 
\end{itemize}\vspace{-1.5em}  %

Together, our results establish the first framework for CRISPR gRNA modeling that leverages pre-trained sequence representations and biologically grounded regulatory features.

\section{Related Works}
\subsection{gRNA activity prediction}

Early models for predicting gRNA activity relied on rule-based scoring or engineered sequence features derived from small datasets \cite{hsu2013dna, xu2015comprehensive, moreno2015crisprscan}, but offered limited generalization. DeepCpf1 \cite{kim2018deep} introduced a CNN-based framework trained on one-hot encoded Cas12a sequences, integrating chromatin accessibility to improve prediction accuracy. Since then, deep learning models have been applied to Cas9 and Cas12 systems, but often remain PAM-specific and trained end-to-end on narrow datasets. Recently, gRNA-FM \cite{zhou2023grnafm} proposed a foundation model tailored for Cas9 guide design, but it does not address Cas12 or activity prediction. Our work fills this gap by leveraging general-purpose biological FMs for Cas12 gRNA activity prediction, independent of PAM motifs and with integrated epigenomic context.

\subsection{Foundation models for DNA and RNA}
 Foundation models (FMs) trained on large-scale biological sequence data have shown strong generalization across a variety of genomic and transcriptomic tasks. FMs such as DNABERT \cite{ji2021dnabert}, DNABERT-2 \cite{zhou2024dnabert}, Evo~\cite{nguyen2024sequence}, Evo-2~\cite{brixi2025genome}, Nucleotide Transformer~\cite{dalla2025nucleotide} are pretrained on genomic DNA, and have been applied to promoter identification, variant effect prediction, and regulatory element classification. Similarly, RNA-FM \cite{chen2022rnafm}, Rinalmo~\cite{penic2024rinalmo}, SpliceBERT~\cite{chen2023self}, UTR-BERT~\cite{yang2024deciphering} and other transcriptomic FMs, capture regulatory and structural features relevant to RNA function and folding. These models generate rich, contextualized embeddings without task-specific supervision. In this work, we adopt RNA-FM and DNABERT-2 as strong, domain-representative FM backbones and evaluate their embeddings for Cas12 gRNA activity prediction.

\section{Methods}

\begin{figure*}[t]
  \centering
  \includegraphics[width=2.\columnwidth]{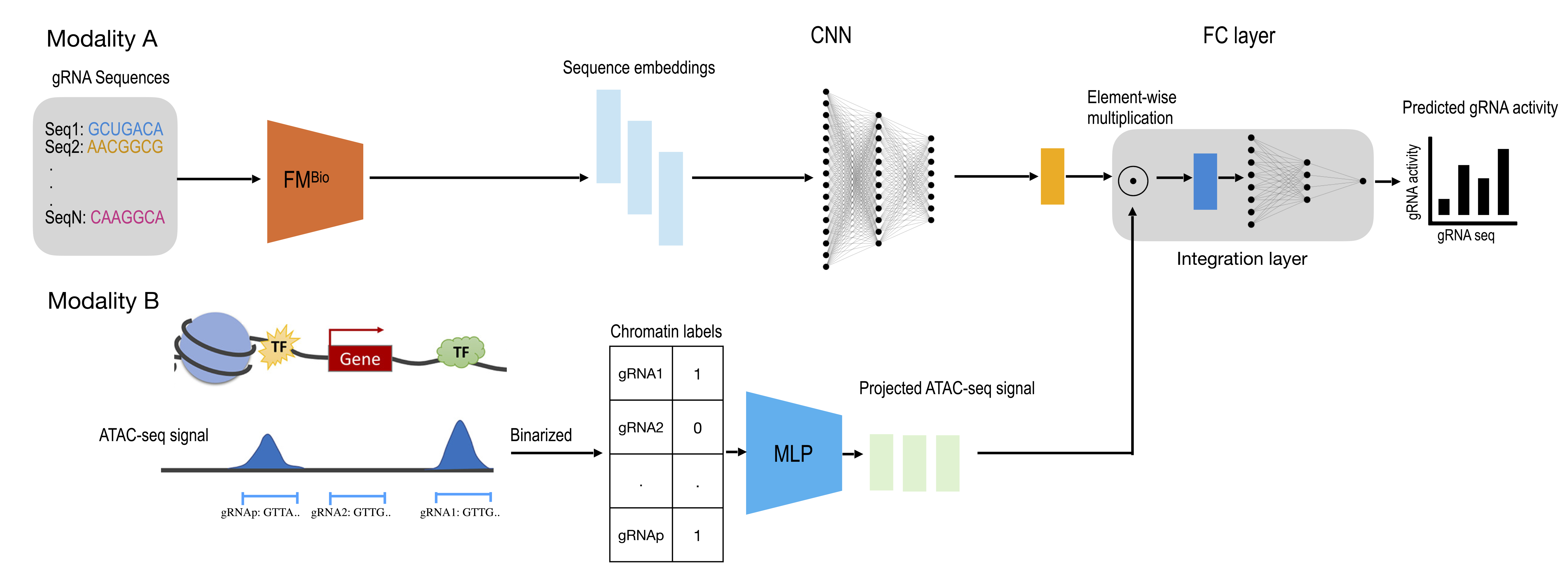}
  \caption{Multimodal framework for predicting CRISPR-Cas12 gRNA activity.
gRNA sequences (Modality A) are encoded using pretrained biological foundation models (FM\textsubscript{Bio}), and the resulting embeddings are processed by a CNN to learn sequence representations. Chromatin accessibility signals (Modality B), derived from binarized ATAC-seq peaks, are projected via an MLP into the same latent space as the sequence features. The two modalities are integrated through an element-wise multiplication layer, followed by a regression head to predict gRNA activity.}
  \label{fig:teaser}
  \vspace{-1em}
\end{figure*}

\subsection{Pretrained Sequence Embeddings}

We model gRNA activity using pre-trained biological FMs to generate contextualized sequence representations. Specifically, we leverage two transformer-based models: \textbf{RNA-FM}~\citep{chen2022rnafm}, trained on transcriptomic sequences, and the \textbf{DNABERT-2}~\citep{zhou2024dnabert}, trained on large genomic DNA corpora owing to their reported SOTA performance in recent studies~\cite{chen2022rnafm, yazdani2024helm, prakash2024bridging, zhou2024dnabert}.

For each gRNA, we extracted not only the core 20-nt target binding sequence but also extended sequence contexts of lengths 34, and 50 nucleotides centered around the cleavage site. This enabled us to systematically investigate how varying amounts of surrounding sequence context influences predictive modeling of gRNA activity.
 These are passed through the frozen FM backbones to obtain fixed-length embeddings, which are used as inputs to a light-weight downstream prediction head. This setup, commonly referred as probing~\cite{prakash2024bridging, lin2023evolutionary} enables the use of rich, pre-trained embeddings while avoiding the need for retraining the base models.

To ensure fair comparison with prior work, we adopt the convolutional regression architecture from~\cite{kim2018deepcpf1}—a state-of-the-art model for Cas12 gRNA activity prediction—as our downstream head, training it on top of pretrained FM embeddings to predict Cas12 activity. This isolates the effect of the FM backbone, as all other components are kept identical to~\cite{kim2018deepcpf1}. The architecture consists of convolutional and pooling layers followed by fully connected layers, terminating in a single scalar output (hyperparameters and training details reported in Appendix sections~\ref{sec:model_arch} and~\ref{sec:model_pram}). This design using FM with CNN prediction head captures local dependencies while remaining efficient for low-data regimes. We refer to this setup—comprising the frozen foundation model backbone paired with the CNN regression head—as Cas-FM (RNA-FM) or Cas-FM (DNABERT-2), depending on the backbone employed.

\subsection{Chromatin Accessibility as an Epigenomic Modality}

To incorporate regulatory context, we integrate chromatin accessibility (CA) information derived from ATAC-seq data (GEO accession: GSM2902624) for the gRNA sequences considered (details of corresponding gRNA data is provided in Sec.~\ref{sec:experiments}). Raw reads were aligned to the hg38 reference genome using \texttt{Bowtie2}~\cite{langmead2012bowtie2}, and peaks were called using \texttt{MACS2}~\cite{zhang2008macs}. We then intersected the resulting peak coordinates with gRNA target loci using \texttt{BEDTools}~\cite{quinlan2010bedtools} to assign binary accessibility labels—gRNAs overlapping a peak with normalized signal $>0.001$ were labeled accessible; others were marked inaccessible.

These binary CA labels were projected into a dense embedding using a multilayer perceptron (MLP), forming the epigenomic input stream. This projected signal was integrated with the sequence-derived feature map from the pretrained FM (backbone) + CNN (downstream head) pipeline via element-wise multiplication in a dedicated integration layer. Keeping the downstream architecture fixed across unimodal (sequence-only) and multimodal (sequence + CA) settings allows us to isolate the contribution of chromatin accessibility to gRNA activity prediction (see Fig.~\ref{fig:teaser}). The exact architecture details are described in Appendix~\ref{sec:model_arch}.

\section{Experiments and results}
\label{sec:experiments}

\begin{figure*}[t]
\centering

\begin{minipage}[t]{0.55\textwidth}
\centering
\footnotesize
\vspace{0pt}  
\begin{tabular}{|l|c|c|l|}
\hline
\textbf{Method} & \textbf{Correlation} & \textbf{CA} & \textbf{Features Used} \\
\hline
CINDEL         & 0.61 & No  & Hand-crafted features \\
Lasso          & 0.64 & No  & One-hot encoding \\
L2             & 0.63 & No  & One-hot encoding \\
L1L2           & 0.64 & No  & One-hot encoding \\
Boosted RT     & 0.66 & No  & One-hot + positional features \\
DeepCpf1       & 0.71 & Yes & One-hot + chromatin accessibility \\
Cas-FM (DNABERT-2)       & 0.49 & No  & Pretrained sequence embedding \\
Cas-FM (RNA-FM)          & \textbf{0.76} & No  & Pretrained RNA embedding \\
Cas-FM-CA (RNA-FM)       & \textbf{0.78} & Yes & RNAFM embedding + CA \\
\hline
\end{tabular}
\captionof{table}{Comparison of gRNA activity prediction methods. Cas-FM (RNA-FM) achieves the highest performance without using chromatic accessibility (CA) modality, while Cas-FM-CA (RNA-FM) further improves performance by incorporating chromatin accessibility as additional modality.}

\label{tab:model_comparison}
\end{minipage}
\hfill
\begin{minipage}[t]{0.25\textwidth}
\centering
\vspace{0.001ex}  
\includegraphics[width=\linewidth]{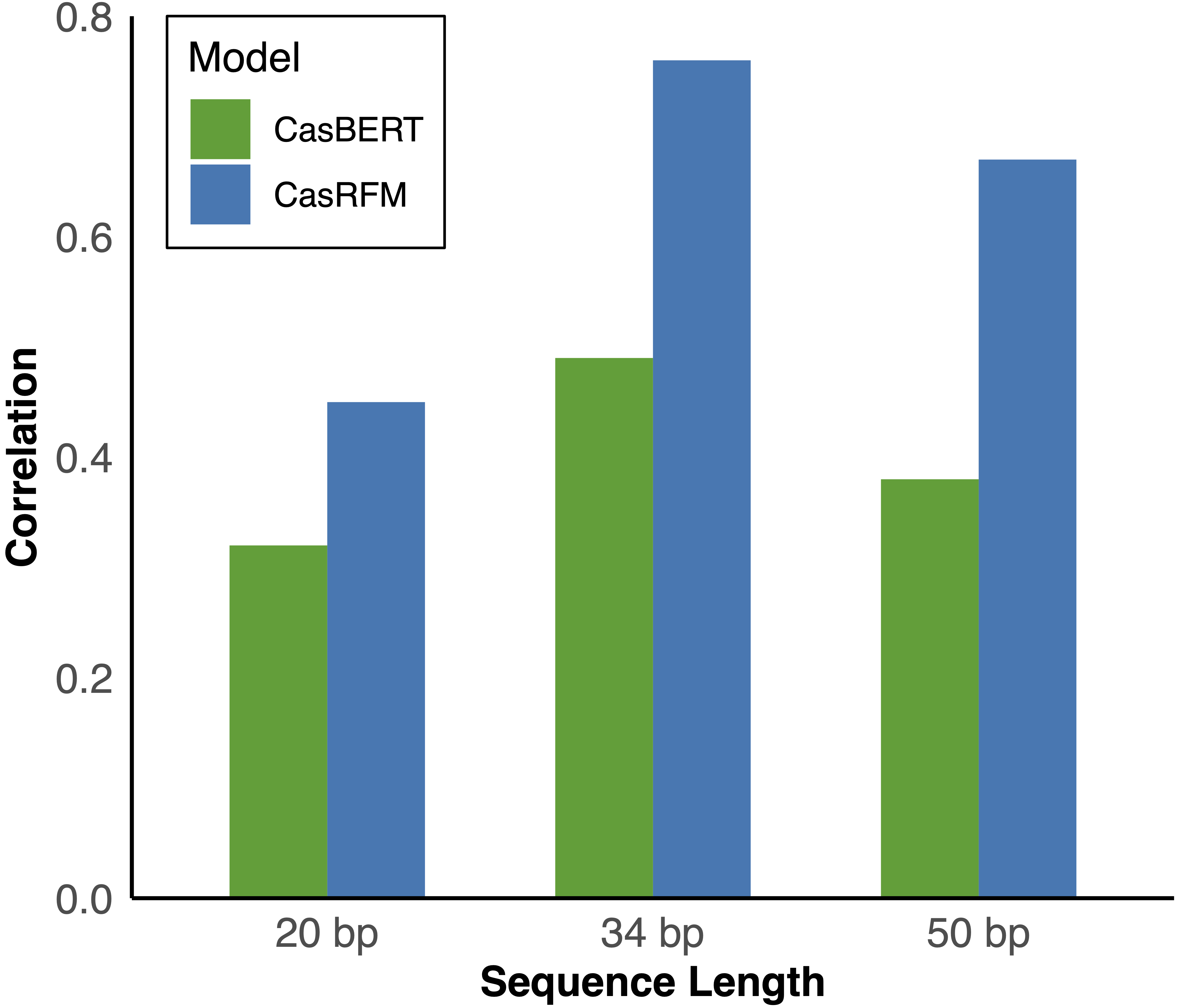}
\captionof{figure}{Performance of Cas-FM variants across varying gRNA sequence lengths, measured by Spearman rank correlation.}
\label{fig:Seqlength}
\end{minipage}

\end{figure*}

\paragraph{Evaluation Dataset.}
We evaluate our approach using the CRISPR-Cas12a (Cpf1) dataset introduced by \citet{kim2018deep}, which contains approximately 15,000 guide RNA (gRNA) sequences paired with experimentally measured cleavage efficiencies. These activity scores reflect the capacity of each guide to induce gene disruption in human cells, making them a biologically grounded target for regression. The dataset spans a range of sequence contexts, including variation in PAM sequences, flanking sequences, and spacer regions, enabling robust model evaluation across heterogeneous genomic backgrounds. More details are provided in Appendix~\ref{sec:appendix_gRNAdataset}. \\

\vspace{-1 em}
\paragraph{Baselines.}
To rigorously benchmark our approach, we compare against a range of baselines commonly used in CRISPR gRNA activity prediction. These include CINDEL~\citep{kim2017vivo}, a logistic regression model originally developed for indel frequency prediction; a boosted regression tree model (Boosted RT) implemented via XGBoost, using k-mer and positional features~\cite{chen2016xgboost}; and three regularized linear regression models—Lasso (L1)\cite{tibshirani1996lasso}, Ridge (L2)\cite{hoerl1970ridge}, and ElasticNet (L1L2)~\cite{zou2005elasticnet}—each trained on curated sets of biologically informed features such as nucleotide composition, GC content, and dinucleotide frequencies.
 These baselines typically use handcrafted sequence features or 1-hot representation of gRNA (see Table~\ref{tab:model_comparison}). Additionally we compared against the state-of-the-art deep learning model DeepCpf1~\cite{kim2018deepcpf1} which has the same CNN architecture as our downstream head and also uses chromatin accessibility as additional modality. We use the same train and test splits as introduced in~\citep{kim2018deep}. The numbers for baselines are directly taken from~\citep{kim2018deep}. 

\paragraph{Pretrained RNA-FM embeddings improve gRNA activity prediction.}

Table~\ref{tab:model_comparison} summarizes the model performance. Cas-FM (RNA-FM)  substantially outperforms all baselines while Cas-FM (DNABERT-2) significantly lags behind. Cas-FM (RNA-FM) works best likely due to the pre-training of RNA-FM backbone on transcriptomic sequences, which closely mirror the biological modality of guide RNAs. These results underscore the advantage of leveraging domain-aligned FMs: pretrained embeddings from RNA-FM capture complex, biologically meaningful dependencies beyond handcrafted features or 1-hot encodings.

\paragraph{Impact of gRNA sequence length/context on activity prediction.} 

We evaluate Cas-FM (RNA-FM) and Cas-FM (DNABERT-2) with three context window sizes centered on the target site: 20 nucleotides (minimal core guide), 34 nucleotides (including flanking and PAM-adjacent bases), and 50 nucleotides (extended upstream and downstream context). Results in Fig~\ref{fig:Seqlength} show that a 34-nt context consistently leads to the best performance.  Cas-FM with either backbone underperforms with shorter 20-nt input, suggesting insufficient contextual information due to the lack of important flanking information that may influence folding, targeting, or interaction with Cas12, while the 50-nt context slightly degrades performance—likely due to the inclusion of less informative or noisy upstream/downstream sequence. These findings highlight the importance of context-aware input design and suggest a moderate-length context window—such as 34 nt—strike a balance between capturing local sequence features relevant to cleavage efficiency (e.g., nucleotide preferences near the PAM site) and avoiding overfitting or dilution from unrelated sequence regions.

\textbf{Impact of chromatin accessibility on gRNA activity prediction}
Table~\ref{tab:model_comparison} also assesses the contribution of epigenomic context for gRNA activity prediction. We evaluated whether incorporating chromatin accessibility (CA) improves the accuracy of gRNA activity prediction.
Driven by the superior performance of Cas-FM with RNA-FM backbone, we limit our experiment with Cas-FM (RNA-FM) alone. Our results show that incorporating chromatin accessibility (denoted as Cas-FM-CA (RNA-FM)) improves predictive performance over using grRNA sequence embeddings alone (Cas-FM (RNA-FM)). This highlights the biological relevance of chromatin state in determining gRNA efficacy. Even when the sequence features are represented by powerful FM embeddings, the addition of CA features captures orthogonal information related to the physical availability of the genomic target site for Cas12-mediated cleavage. In practical terms, guides targeting open chromatin regions are more likely to be effective, and our model is better able to reflect that when explicitly given this contextual information.

These findings underscore the value of integrating multi-modal biological data, combining large-scale pretrained sequence representations with relevant experimental measurements, to more accurately model the determinants of CRISPR guide activity.

\section{Discussion}
Our study demonstrates that the pre-trained transcriptomic foundational model RNA-FM significantly enhances the prediction of CRISPR-Cas12 gRNA activity compared to existing methods, even in out-of-distribution settings with short gRNA sequences that the foundation model was not pre-trained on. This is likely attributable to the fact that, from a biological perspective, gRNA sequences are functionally embedded within broader transcriptomic contexts. Their activity relies on factors such as sequence composition, secondary structure, and chromatin accessibility, which are partially captured by RNA-FM.

While prior work has noted the importance of incorporating extended sequence context around the gRNA~\citep{kim2018deepcpf1}, our findings reinforce this insight in the context of foundation model-based learning. We confirm that a 34-nucleotide context strikes the best balance between informativeness and generalization, outperforming both shorter and longer input windows across different model variants.

Finally, we introduce a curated dataset of demonstrate of chromatin accessibility features and show that integrating these in a multi-modal setting with gRNA sequences further enhances predictive accuracy. This suggests that epigenomic context introduces complementary information not captured by sequence alone, such as physical accessibility of the target site. Taken together, our work points toward a powerful direction for CRISPR modeling: leveraging domain-aware FMs while incorporating biologically grounded, orthogonal signals to better predict genome editing outcomes.


\nocite{langley00}

\bibliography{example_paper}
\bibliographystyle{icml2025}

\newpage
\appendix
\onecolumn
\section{ Appendix}
\subsection{gRNA Activity Dataset }
\label{sec:appendix_gRNAdataset}

The gRNA activity dataset used in this study originates from the HT1 high-throughput screening experiment described in the DeepCpf1 paper \cite{kim2018deepcpf1}. In this experiment, synthetic gRNA-target constructs were integrated into the genome of HEK293T cells using lentiviral delivery. Each construct encoded a 20-nt guide RNA sequence, a unique barcode, and a matching target site, allowing systematic evaluation of Cpf1 cleavage activity. The activity of each gRNA was quantified as the indel frequency measured via deep sequencing, corrected by background subtraction. For each guide, the dataset includes multiple representations of the target context: the core 20-nt gRNA sequence, an extended 34-bp target sequence (including PAM and flanking regions), and a 50-bp sequence window capturing broader genomic context around the cleavage site. This design supports modeling across multiple levels of sequence context.

Following the original data preparation strategy, we partitioned the dataset into two subsets: HT1-1 ($n= 15000$), which was used for model training, validation, and HT1-2 ($n = 1290$) for testing.

\subsection{ATAC-seq data curation }
\label{sec:appendix_ATACdataset}

\begin{figure}[H]
  \centering
  \includegraphics[width=\textwidth]{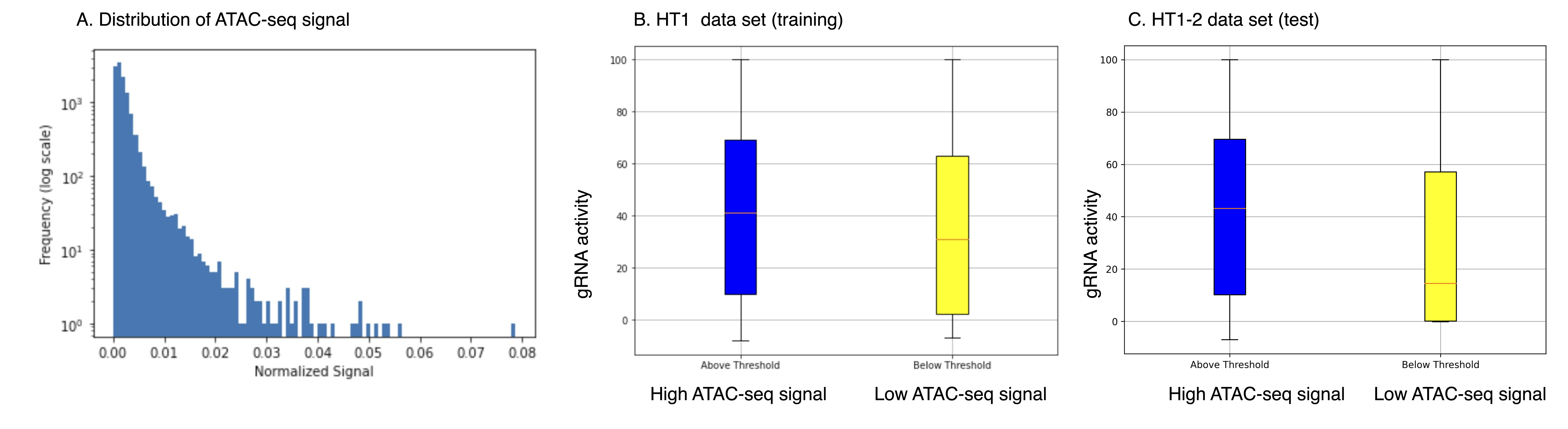}
  \caption{Association of continuous ATAC-seq signal with gRNA activity. A) Genome wide distribution of ATAC-seq signal in \textit{HEK293T} cell line. B-C) ATAC-seq signal intensity is correlated with gRNA activity both in training and test data }
  \label{fig:gRNAActivity}
\end{figure}

To annotate gRNA target sites with chromatin accessibility (CA) status, we processed publicly available ATAC-seq data for the \textit{HEK293T} cell line obtained from the GEO database (accession: GSM2902624). We specifically used the  FASTQ files: \texttt{HEK293\_ATAC\_medium\_depth\_bio2\_tech1\_1.fastq.gz} and \texttt{HEK293\_ATAC\_medium\_depth\_bio2\_tech1\_2.fastq.gz}.

The raw data were preprocessed through a standard ATAC-seq pipeline. First, \textit{alignment} of the reads to the human reference genome (\textit{hg38}) was performed using \texttt{Bowtie2}~~\cite{langmead2012bowtie2} with default parameters. Following alignment, \textit{peak calling} was performed using \texttt{MACS2}~~\cite{zhang2008macs} to identify regions of open chromatin, reported as narrow peaks. Next, we performed normalization and thresholding of the \texttt{MACS2} signal intensities, applying a conservative cut-off of \$0.001\$ . Genomic regions with a normalized signal above this threshold were considered accessible, while those below were treated as inaccessible.

To annotate each gRNA with a chromatin accessibility (CA) label, we first \textit{aligned} the gRNA sequences to the human reference genome (\textit{hg38}) using \texttt{BLAT}~~\cite{kent2002blat}, a fast alignment tool suitable for short nucleotide queries. For each gRNA, we extracted the best-matching locus and recorded the corresponding genomic coordinates (chromosome, start, end, and strand). These coordinates were then converted into standard \texttt{BED} format to facilitate downstream processing. We \textit{intersected} the gRNA BED file with ATAC-seq peak regions using \texttt{bedtools intersect}~~\cite{quinlan2010bedtools}, and assigned an accessibility label of $1$ (accessible) if a gRNA overlapped with a peak exceeding the signal threshold of $0.001$, or $0$ (inaccessible) otherwise. This resulted in a binary CA label for each gRNA, which

The raw data were preprocessed through a standard ATAC-seq pipeline. First, \textit{alignment} of the reads to the human reference genome (\textit{hg38}) was performed using \texttt{Bowtie2} with default parameters. Following alignment, \textit{peak calling} was performed using \texttt{MACS2} to identify regions of open chromatin, reported as narrow peaks. Next, we performed normalization and thresholding of the \texttt{MACS2} signal intensities, applying a conservative cut-off of $0.001$ . Genomic regions with a normalized signal above this threshold were considered accessible, while those below were treated as inaccessible. This resulted in a binary CA label for each gRNA, which was used as input feature in the downstream modeling and analysis.

We partitioned the genome into \textit{highly accessible} and \textit{low-accessibility} regions based on chromatin accessibility scores, using a threshold of 0.001 on the normalized ATAC-seq signal. This threshold was applied consistently across both training and test datasets to classify each gRNA target site as accessible or inaccessible.

To assess how accessibility influences gRNA efficacy, we compared the distribution of experimentally measured gRNA activity scores across the two accessibility categories. As shown in Figure~\ref{fig:gRNAActivity}, gRNAs targeting accessible regions exhibited significantly higher activity than those in inaccessible regions. This observation suggests a positive correlation between chromatin openness and Cpf1 cleavage efficiency, consistent with the biological intuition that accessible chromatin is more permissive to endonuclease activity.

\subsection{Model architecture}
\label{sec:model_arch}

\begin{algorithm}
\dirtree{%
.0 .
.1 Input modality.
.2 gRNA embedding \dotfill \{RNA-FM T12 (640-dim)\}.
.2 Chromatin accessibility label \dotfill \{Binary label (1-dim)\}.
.1 CNN layer.
.2 Conv1D \dotfill \{kernel=5, filters=80, activation=ReLU\}.
.2 AvgPool1D \dotfill \{pool size=2\}.
.2 Flatten.
.2 Dropout \dotfill \{rate=0.3\}.
.2 Dense layer \dotfill \{units=80, activation=ReLU\}.
.2 Dropout \dotfill \{rate=0.3\}.
.2 Dense layer \dotfill \{units=40, activation=ReLU\}.
.2 Dropout \dotfill \{rate=0.3\}.
.2 Dense layer \dotfill \{units=40, activation=ReLU\}.
.1 Chromatin encoder.
.2 Dense layer \dotfill \{units=40, activation=ReLU\}.
.1 Integration layer.
.2 Element-wise multiplication \dotfill \{between gRNA sequence and chromatin accessibility\}.
.1 Output head.
.2 Dropout \dotfill \{rate=0.3\}.
.2 Dense layer \dotfill \{units=1, activation=linear\}.
}
\caption{Cas-FM (RNA-FM) architecture. \label{alg:deepcpf1_architecture}}
\end{algorithm}

\subsection{Model training hyperparameter}
\label{sec:model_pram}

We trained the model for $100$ epochs using the Adam optimizer with a learning rate of $5 \times 10^{-5}$ and a batch size of $32$. Model selection was based on validation loss, with early stopping applied after $10$ epochs of no improvement. Mean Squared Error (MSE) was used as the training objective for the regression task, and Spearman rank correlation was computed on the test set for evaluation. Learning rate scheduling was performed using ReduceLROnPlateau with a patience of $5$ epochs and a decay factor of $0.1$.


\end{document}